\documentstyle[12pt]{article}
\begin{document}

\thispagestyle{empty}
\renewcommand{\thefootnote}{\fnsymbol{footnote}}
\setcounter{footnote}{0}

\def\pxb{\left(\p \times \B - \B \times \p \right)}
\def\LAMBDA{\bar{\lambda}}
\def\rk{r_k}
\def\beq{\begin{equation}}
\def\eeq{\end{equation}}
\def\bea{\begin{eqnarray}}
\def\eea{\end{eqnarray}}
\def\nn{\nonumber}
\def\ba{\begin{array}}
\def\ea{\end{array}}
\def\0{{\mbox{\boldmath $0$}}}
\def\one{1\hskip -1mm{\rm l}}
\def\A{{\mbox{\boldmath $A$}}}
\def\B{{\mbox{\boldmath $B$}}}
\def\El{{\mbox{\boldmath $E$}}}
\def\F{{\mbox{\boldmath $F$}}}
\def\S{{\mbox{\boldmath $S$}}}

\def\a{{\mbox{\boldmath $a$}}}
\def\p{{\mbox{\boldmath $p$}}}
\def\hatp{{\hat{\mbox{\boldmath $p$}}}}
\def\hatP{{\hat{\mbox{\boldmath $P$}}}}
\def\vpi{{\mbox{\boldmath $\pi$}}}
\def\hatvpi{\hat{\mbox{\boldmath $\pi$}}}
\def\r{{\mbox{\boldmath $r$}}}
\def\v{{\mbox{\boldmath $v$}}}
\def\w{{\mbox{\boldmath $w$}}}
\def\H{{\rm H}}
\def\hA{\hat{A}}
\def\hB{\hat{B}}
\def\i{{\rm i}}
\def\ih{\frac{\i}{\hbar}}
\def\ixh{\i \hbar}
\def\ddx{\frac{\partial}{\partial x}}
\def\ddy{\frac{\partial}{\partial y}}
\def\ddz{\frac{\partial}{\partial z}}
\def\ddt{\frac{\partial}{\partial t}}
\def\vsig{{\mbox{\boldmath $\sigma$}}}
\def\Al{{\mbox{\boldmath $\alpha$}}}
\def\ho{\hat{\cal H}_o}
\def\half{\frac{1}{2}}
\def\E{{\hat{\cal E}}}
\def\O{{\hat{\cal O}}}
\def\eps{\epsilon}
\def\g{\gamma}
\def\Vomeg{{\underline{\mbox{\boldmath $\Omega$}}}_s}
\def\hH{\hat{H}}
\def\Vsig{{\mbox{\boldmath $\Sigma$}}}
\def\Nab{{\mbox{\boldmath $\nabla$}}}
\def\curl{{\rm curl}}
\def\bh{\bar{H}}
\def\th{\tilde{H}}

\def\At{{\hat{A}(t)}}
\def\dAt{\frac{\partial {\hat{A}(t)} }{\partial t}}
\def\sone{{\hat{S}_1}}
\def\dsone{\frac{\partial {\hat{S}_1} }{\partial t}}
\def\dO{\frac{\partial {\hat{\cal O}}}{\partial t}}

\begin{center}
{\Large\bf
Maxwell Optics: III. Applications}

\bigskip

{\em Sameen Ahmed KHAN} \\

\bigskip

khan@fis.unam.mx ~~~ http://www.pd.infn.it/$\sim$khan/ \\
rohelakhan@yahoo.com ~~~ http://www.imsc.ernet.in/~jagan/khan-cv.html \\
Centro de Ciencias F\'{i}sicas,
Universidad Nacional Aut\'onoma de M\'exico, \\
Apartado Postal 48-3,
Cuernavaca 62251,
Morelos, \\
{\bf M\'EXICO} \\

\end{center}

\bigskip

\noindent
{\bf Abstract} \\
A new formalism of beam-optics and polarization has been recently
presented, based on an exact matrix representation of the Maxwell
equations.  This is described in Part-I and Part-II.  In this Part,
we present the application of the above formalism to the specific
example of the axially symmetric graded index fiber.  This formalism
leads to the wavelength-dependent modifications of the six
aberrations present in the traditional prescriptions and further gives
rise to the remaining three aberrations permitted by the axial
symmetry.  Besides, it also gives rise to a wavelength-dependent
image rotation.  The three extra aberrations and the image rotation
are not found in any of the traditional approaches.

\bigskip

\section{Introduction}
In Part-I and Part-II we presented the exact matrix representation of
the Maxwell equations in a medium with varying permittivity and
permeability~\cite{Khan-2,Khan-3}.  From this we derived an exact
optical Hamiltonian, which was shown to be in close algebraic analogy
with the Dirac equation.  This enabled us to apply the machinery of the
Foldy-Wouthuysen transformation and we obtained an expansion for the
beam-optical Hamiltonian which works to all orders.  Formal expressions
were obtained for the paraxial Hamiltonian and the leading order
aberrating Hamiltonian, without assuming any form for the refractive
index.  Even at the paraxial level the wavelength-dependent effects
manifest by the presence of a matrix term coupled to the logarithmic
gradient of the refractive index.  This matrix term is very similar
to the spin term in the Dirac equation and we call it as
the {\em polarizing term} in our formalism.  The aberrating Hamiltonian
contains numerous wavelength-dependent terms in two guises:  One of
these is the explicit wavelength-dependent terms coming from the
commutators inbuilt in the formalism with $\LAMBDA$ playing the role
played by $\hbar$ in quantum mechanics.  The other set arises from the
the polarizing term.

Now, we apply the formalism to specific examples.  One is the medium
with constant refractive index.  This is perhaps the only problem
which can be solved exactly in a closed form expression.  This is just
to illustrate how the aberration expansion in our formalism can be
summed to give the familiar exact result.

The next example is that of the axially symmetric graded index medium.
This example enables us to demonstrate the power of the formalism,
reproducing the familiar results from the traditional approaches and
further giving rise to new results, dependent on the wavelength.

\section{Medium with Constant Refractive Index}
Constant refractive index is the simplest possible system.  In our
formalism, this is perhaps the only case where it is possible to do
an exact diagonalization.  This is very similar to the exact
diagonalization of the free Dirac Hamiltonian.  From the experience of
the Dirac theory we know that there are hardly any situations where one
can do the exact diagonalization.  One necessarily has to resort to
some approximate diagonalization procedure.  The Foldy-Wouthuysen
transformation scheme provides the most convenient and accurate
diagonalization to any desired degree of accuracy.
So we have adopted the Foldy-Wouthuysen scheme in our formalism.

For a medium with constant refractive index,
$n \left(\r \right) = n_c$, we have,
\bea
\hat{\H}_{c}
& = & - n_c \beta +
\i \left(M_y p_x - M_x p_y \right)\,,
\label{H-Constant}
\eea
which is exactly diagonalized by the following transform,
\bea
T^{\pm}
& = &
\exp{\left[\i \left(\pm \i \beta \right) \O \theta \right]} \nn \\
& = &
\exp{ \left[\mp \i \beta
\left(M_y p_x - M_x p_y \right)
\theta \right]} \nn \\
& = &
\cosh \left(\left|\hatp_{\perp} \right| \theta \right)
\mp \i \frac{\beta
\left(M_y p_x - M_x p_y \right)}
{\left|\hatp_{\perp}\right|}
\sinh \left(\left|\hatp_{\perp} \right| \theta \right)
\label{T-D}
\eea
We choose,
\bea
\tanh \left(2 \left|\hatp_{\perp} \right| \theta \right)
=
\frac{\left|\hatp_{\perp} \right|}{n_c}
\label{tanh}
\eea
then
\bea
T^{\pm}
=
\frac{\left(n_c + P_z \right)
\mp \i \beta \left(M_y p_x - M_x p_y \right)}
{\sqrt{2 P_z \left(n_c + P_z \right)}}
\eea
where $P_z = + \sqrt{\left(n_c^{2} - \hatp_{\perp}^2 \right)}$.
Then we obtain,
\bea
\hat{\H}_c^{\rm diagonal}\,
& = &
T^{+} \hat{\H}_{c} T^{-} \nn \\
& = &
T^{+} \left\{- n_c \beta
+ \i \left(M_y p_x - M_x p_y \right) \right\} T^{-} \nn \\
& = &
- \left\{n_c^{2} - \hatp_{\perp}^2 \right\}^{\frac{1}{2}} \beta
\label{Constant-Diagonal}
\eea
We next, compare the exact result thus obtained with the approximate
ones, obtained through the systematic series procedure we have developed.
\bea
\hat{\cal H}^{(4)}_c
& = &
- n_c
\left\{1 - \frac{1}{2 n_c^2} \hatp_{\perp}^2
- \frac{1}{8 n_c^4} \hatp_\perp^4 - \cdots \right\} \beta \nn \\
& \approx &
- n_c \left\{1 - \frac{1}{n_c^2} \hatp_\perp^2\right\}^{\frac{1}{2}} \beta \nn \\
& = &
- \left\{n_c^2 - \hatp_\perp^2 \right\}^{\frac{1}{2}} \beta \nn \\
& = &
\hat{\H}_c^{\rm diagonal}\,.
\label{Constant-approximate}
\eea

Knowing the Hamiltonian, we can compute the transfer maps.  The
transfer operator between any pair of points
$\left\{(z^{\prime \prime} , z^{\prime}) \left|
z^{\prime \prime} \right. > z^{\prime} \right\}$
on the $z$-axis, is formally given by
\bea
\left|\psi (z^{\prime \prime} , z^{\prime}) \right|
=
\hat{\cal T} (z^{\prime \prime} , z^{\prime})
\left|\psi (z^{\prime \prime} , z^{\prime}) \right\rangle \,,
\label{}
\eea
with
\bea
& & \i \LAMBDA \frac{\partial}{\partial z}
\hat{\cal T} (z^{\prime \prime} , z^{\prime})
=
\hat{\cal H} \hat{\cal T} (z^{\prime \prime} , z^{\prime})\,, \quad
\hat{\cal T} (z^{\prime \prime} , z^{\prime})
=
\hat{\cal I}\,, \nn \\
& & \nn \\
& & \hat{\cal T} (z^{\prime \prime} , z^{\prime})
=
\wp \left\{\exp
\left[- \frac{\i}{\LAMBDA}
\int_{z^\prime}^{z^{\prime \prime}} dz\,
\hat{\cal H} (z) \right] \right\} \nn \\
& & \quad
=
\hat{\cal I}
- \frac{\i}{\LAMBDA} \int_{z^\prime}^{z^{\prime \prime}} dz
\hat{\cal H} (z) \nn \\
& & \qquad
+ \left(- \frac{\i}{\LAMBDA}\right)^2
\int_{z^\prime}^{z^{\prime \prime}} dz
\int_{z^\prime}^{z} d z^\prime
\hat{\cal H} (z) \hat{\cal H} (z^\prime) \nn \\
& & \qquad
+ \ldots\,,
\label{Transfer-1}
\eea
where $\hat{\cal I}$ is the identity operator and $\wp$ denotes the
path-ordered exponential.  There is no closed form expression for
$\hat{\cal T} (z^{\prime \prime} , z^{\prime})$ for an arbitrary choice
of the refractive index $n (\r)$.  In such a situation the most
convenient form of the expression for the $z$-evolution
operator $\hat{\cal T} (z^{\prime \prime} , z^{\prime})$, or the
$z$-propagator, is
\beq
\hat{\cal T} (z^{\prime \prime} , z^{\prime})
=
\exp{
\left[- \frac{\i}{\LAMBDA} \hat{T}
(z^{\prime \prime} , z^{\prime}) \right]}\,,
\label{Transfer-2}
\eeq
with
\bea
\hat {T} (z^{\prime \prime} , z^{\prime})
& = &
\int_{z^\prime}^{z^{\prime \prime}} dz \hat{\cal H} (z) \nn \\
& & \qquad
+ \frac{1}{2} \left(- \frac{\i}{\LAMBDA} \right)
\int_{z^\prime}^{z^{\prime \prime}} dz
\int_{z^\prime}^{z}  d z^\prime
\left[\hat{\cal H} (z)\,, \hat{\cal H} (z^\prime) \right] \nn \\
& & \qquad
+ \ldots \,,
\label{T-Magnus}
\eea
as given by the Magnus formula~\cite{Magnus}.  We shall be needing
these expressions in the next example where the refractive index is
not a constant.

Using the procedure outlined above we compute the transfer operator,
\bea
& & \hat{U}_c \left(z_{\rm out}, z_{\rm in} \right)
= \exp{ \left[- \frac{\i}{\LAMBDA} \Delta z {\cal H}_c \right]} \nn \\
& & =
\exp{ \left[+ \frac{\i}{\LAMBDA} n_c \Delta z
\left\{1 - \frac{1}{2} \frac{\hat{p}_{\perp}^2}{n_c^2}
- \frac{1}{8} \left(\frac{\hat{p}_{\perp}^2}{n_c^2} \right)^2
- \cdots \right\} \right]}\,, 
\label{Constant-U}
\eea
where, $\Delta z = \left(z_{\rm out}, z_{\rm in} \right)$.
Using~(\ref{Constant-U}), we compute the transfer maps
\bea
\left(
\ba{c}
\left\langle \r_{\perp} \right\rangle \\
\left\langle \p_{\perp} \right\rangle
\ea
\right)_{\rm out}
=
\left(
\ba{ccc}
1 & & \frac{1}{\sqrt{n_c^2 - \p_{\perp}^2}} \Delta z \\
0 & & 1
\ea
\right)
\left(
\ba{c}
\left\langle \r_{\perp} \right\rangle \\
\left\langle \p_{\perp} \right\rangle
\ea
\right)_{\rm in}\,.
\label{Constant-Maps}
\eea
The beam-optical Hamiltonian is intrinsically aberrating.  Even for
simplest situation of a constant refractive index, we have aberrations
to all orders.

\section{Axially Symmetric Graded Index Medium}

The refractive index of an axially symmetric graded-index
material can be most generally described by the following
polynomial (see, pp. 117 in~\cite{DFW})
\bea
n \left(\r \right) = n_0 + \alpha_2 (z) \r_{\perp}^2
+ \alpha_4 (z) \r_{\perp}^4 + \cdots\,,
\label{n-Dragt}
\eea
where, we have assumed the axis of symmetry to coincide
with the optic-axis, namely the $z$-axis without any
loss of generality.
We note,
\bea
\hat{\cal E}
& = &
- \left\{
\alpha_2 (z) \r_{\perp}^2
+ \alpha_4 (z) \r_{\perp}^4 + \cdots\,, \right\} \beta
- \i \LAMBDA \beta
{\mbox{\boldmath $\Sigma$}} \cdot {\mbox{\boldmath $u$}} \nn \\
\hat{\cal O}
& = &
\i \left(M_y p_x - M_x p_y \right) \nn \\
& = &
\beta \left({\mbox{\boldmath $M$}}_{\perp} \cdot \hatp_{\perp} \right)
\eea
where
\bea
{\mbox{\boldmath $\Sigma$}} \cdot {\mbox{\boldmath $u$}}
& = &
- \frac{1}{n_0} \alpha_2 (z)
{\mbox{\boldmath $\Sigma$}}_{\perp} \cdot \r_{\perp}
- \frac{1}{2 n_0}
\left(\frac{d }{d z} \alpha_2 (z)\right) \Sigma_z \r_\perp^2
\eea

To simplify the formal expression for the beam-optical Hamiltonian
$\hat{\cal H}^{(4)}$ given in~(24-25) in Part-II, we make use of the
following:
\bea
\left({\mbox{\boldmath $M$}}_{\perp} \cdot \hatp_{\perp} \right)^2
& = &
\hatp_{\perp}^2\,, \quad \quad
\O^2
=
- \hatp_{\perp}^2\,, \quad \quad
\ddz \O = 0\,, \nn \\
\left({\mbox{\boldmath $M$}}_{\perp} \cdot \hatp_{\perp} \right)
\r_{\perp}^2
\left({\mbox{\boldmath $M$}}_{\perp} \cdot \hatp_{\perp} \right)
& = &
\frac{1}{2}
\left(\r_{\perp}^2 \hatp_{\perp}^2
+ \hatp_{\perp}^2 \r_{\perp}^2 \right)
+ 2 \LAMBDA \beta \hat{L}_z
+ 2 \LAMBDA^2\,,
\eea
where, $\hat{L}_z$ is the angular momentum.
Finally, the beam-optical Hamiltonian to order
$\left(\frac{1}{n_0^2} \hatp_\perp^2\right)^2$ is
\bea
\hat{\cal H}
& = &
\hat{H}_{0\,, p}
+ \hat{H}_{0\,, (4)}
+ \hat{H}_{0\,, (2)}^{(\LAMBDA)}
+ \hat{H}_{0\,, (4)}^{(\LAMBDA)}
+ \hat{H}^{(\LAMBDA, \sigma)} \nn \\
\hat{H}_{0\,, p}
& = &
- n_0 + \frac{1}{2 n_0} \hatp_{\perp}^2
- \alpha_2 (z) \r_{\perp}^2 \nn \\
\hat{H}_{0\,, (4)}
& = &
\frac{1}{8 n_0^3} \hatp_{\perp}^4 \nn \\
& &
- \frac{\alpha_2 (z)}{4 n_0^2} \left(\r_{\perp}^2 \hatp_{\perp}^2
+ \hatp_{\perp}^2 \r_{\perp}^2 \right) \nn \\
& &
- \alpha_4 (z) \r_{\perp}^4 \nn \\
\hat{H}_{0\,, (2)}^{(\LAMBDA)}
& = &
- \frac{\LAMBDA^2}{2 n_0^2} \alpha_2 (z)
- \frac{\LAMBDA}{2 n_0^2} \alpha_2 (z) \hat{L}_z
+ \frac{\LAMBDA^2}{2 n_0^3} \alpha_2^2 (z) \r_{\perp}^2 \nn \\
\hat{H}_{0\,, (4)}^{(\LAMBDA)}
& = &
\frac{\LAMBDA}{4 n_0^3} \alpha_2^2 (z)
\left(\r_{\perp}^2 \hat{L}_z
+ \hat{L}_z \r_{\perp}^2 \right)
+
\frac{\LAMBDA^2}{2 n_0^3} \alpha_2 (z) \alpha_4 (z) \r_\perp^4 \nn \\
\hat{H}^{(\LAMBDA, \sigma)}
& = &
\frac{\i \LAMBDA^3}{2 n_0^3}
\left\{\frac{d }{d z} \alpha_2 (z) \right\}
\beta \Sigma_z \nn \\
& & \qquad
+
\frac{\i \LAMBDA^2}{4 n_0^3}
\alpha_2 (z)
\left(\Sigma_x p_y - \Sigma_y p_x \right) \nn \\
& & \qquad
+
\frac{\i \LAMBDA^3}{2 n_0^3}
\left\{\frac{d }{d z} \alpha_2 (z) \right\}
\Sigma_z \hat{L}_z  \nn \\
& & \qquad
+ \frac{\i \LAMBDA}{4 n_0^3}
\alpha_2 (z)
\beta
\left[
{\mbox{\boldmath $\Sigma$}}_\perp \cdot \r_\perp ,
\hatp_{\perp}^2 \right]_{+} \nn \\
& & \qquad
+ \frac{\i \LAMBDA}{8 n_0^3}
\left\{\frac{d }{d z} \alpha_2 (z) \right\}
\beta \Sigma_z
\left[\r_\perp^2 , \hatp_{\perp}^2 \right]_{+} \nn \\
& & \qquad
+ \cdots
\label{H-Fiber}
\eea
where $[A , B] = (AB + BA)$ and `$\cdots$' are the numerous other
terms arising from the polarization term.  We have retained only the
leading order of such terms above for an illustration.  All these
matrix terms, related to the polarization will be addressed elsewhere.

The reasons for partitioning the beam-optical Hamiltonian
$\hat{\cal H}$ in the above manner are as follows.  The paraxial
Hamiltonian, $\hat{H}_{0\,, p}$, describes the ideal behaviour.
$\hat{H}_{0\,, (4)}$ is responsible for the third-order aberrations.
 Both of these Hamiltonians are modified by the wavelength-dependent
contributions given in $\hat{H}_{0\,, (2)}^{(\LAMBDA)}$ and
$\hat{H}_{0\,, (4)}^{(\LAMBDA)}$ respectively.  Lastly, we have
$\hat{H}^{(\LAMBDA, \sigma)}$, which is associated with the
polarization.

From these sub-Hamiltonians we make several observations:
The term
$\frac{\LAMBDA}{2 n_0^2} \alpha_2 (z) \hat{L}_z$ which
contributes to the paraxial Hamiltonian, gives rise to
an {\em image rotation} by an angle $\theta (z)$:
\bea
\theta (z^{\prime \prime} , z^{\prime})
=
\frac{\LAMBDA}{2 n_0^2}
\int_{z^\prime}^{z^{\prime \prime}} d z \alpha_2 (z)\,.
\label{theta}
\eea
This image rotation (which need not be small) has no
analogue in the {\em square-root approach}~\cite{DFW,Dragt-Wave}
and the {\em scalar approach}~\cite{KJS-1,Khan-1}.

The Hamiltonian $\hat{H}_{0\,, (4)}$ is the one we have in the
traditional prescriptions and is responsible for the six aberrations.
$\hat{H}_{0\,, (4)}^{(\LAMBDA)}$ modifies the above six aberrations
by wavelength-dependent contributions and further gives rise to the
remaining three aberrations permitted by the axial symmetry.
Before proceeding further we enumerate all the nine
aberrations permitted by the axial symmetry.
The axial symmetry permits {\em exactly} nine third-order
aberrations which are:

\bigskip

\begin{tabular}{lll}
Symbol & Polynomial & Name \\
$C$ & $\hatp_{\perp}^4$ & Spherical Aberration \\
$K$ & $\left[\hatp_{\perp}^2 \,, \left(\hatp_\perp \cdot \r_\perp
+ \r_\perp \cdot \hatp_\perp \right) \right]_{+}$
& Coma \\
$k$ & $\hatp_{\perp}^2 \hat{L}_z$ & Anisotropic Coma \\
$A$ & $\left(\hatp_\perp \cdot \r_\perp
+ \r_\perp \cdot \hatp_\perp \right)^{2}$ & Astigmatism \\
$a$ & $\left(\hatp_\perp \cdot \r_\perp
+ \r_\perp \cdot \hatp_\perp \right) \hat{L}_z$
& Anisotropic Astigmatism \\
$F$ & $\left(\hatp_{\perp}^2 \r_{\perp}^2
+ \r_{\perp}^2 \hatp_{\perp}^2 \right)$ & Curvature of Field \\
$D$ & $\left[\r_{\perp}^2 \,, \left(\hatp_\perp \cdot \r_\perp
+ \r_\perp \cdot \hatp_\perp \right) \right]_{+}$
& Distortion \\
$d$ & $\r_{\perp}^2 \hat{L}_z$ & Anisotropic Distortion \\
$E$ & $\r_{\perp}^4$ & Nameless? or POCUS
\end{tabular}

~\\
The name {\it POCUS} is used in~\cite{DFW} on page~137.

The axial symmetry allows only the terms (in the Hamiltonian)
which are produced out of,
$\hatp_{\perp}^2$,
$\r_{\perp}^2$,
$\left(\hatp_\perp \cdot \r_\perp + \r_\perp \cdot \hatp_\perp \right)$
and
$\hat{L}_z$.
Combinatorially, to fourth-order one would get ten terms including
$\hat{L}_z^2$.  We have listed nine of them in the table above.  The
tenth one namely,
\bea
\hat{L}_z^2
=
\frac{1}{2}
\left(\hatp_{\perp}^2 \r_{\perp}^2
+ \r_{\perp}^2 \hatp_{\perp}^2 \right)
-
\frac{1}{4}
\left(\hatp_\perp \cdot \r_\perp
+ \r_\perp \cdot \hatp_\perp \right)^{2}
+
\LAMBDA^2
\eea
So, $\hat{L}_z^2$ is not listed separately.  Hence, we have only nine
third-order aberrations permitted by axial symmetry, as stated earlier.

The paraxial transfer maps are given by
\bea
\left(
\ba{c}
\left\langle \r_{\perp} \right\rangle \\
\left\langle \p_{\perp} \right\rangle
\ea
\right)_{\rm out}
=
\left(
\ba{cc}
P & Q \\
R & S
\ea
\right)
\left(
\ba{c}
\left\langle \r_{\perp} \right\rangle \\
\left\langle \p_{\perp} \right\rangle
\ea
\right)_{\rm in}\,,
\label{Paraxial-Maps}
\eea
where~$P$, $Q$, $R$ and $S$ are the solutions of the paraxial
Hamiltonian~(\ref{H-Fiber}). The symplecticity condition tells
us that $P S - Q R = 1$.  In this particular case from the
structure of the paraxial equations we can further conclude
that: $R = P^\prime$ and $S = Q^\prime$ where $^\prime$ denotes
the $z$-derivative.

The transfer operator is most accurately expressed in terms of the
paraxial solutions, $P$, $Q$, $R$ and $S$, {\em via} the
{\em interaction picture}~\cite{Interaction}.
\bea
\hat{\cal T} \left(z\,, z_0\right)
& = &
\exp {
\left[
- \frac{\i}{\LAMBDA} \hat{T}
\left(z\,, z_0\right) \right] }\,, \nn \\
& = &
\exp
\left[
- \frac{\i}{\LAMBDA}
\left\{
C \left(z^{\prime \prime}\,, z^\prime \right) \hatp_{\perp}^4
\phantom{\frac{\i}{\LAMBDA}} \right. \right. \nn \\
& & \quad \quad
+
K \left(z^{\prime \prime}\,, z^\prime \right)
\left[\hatp_{\perp}^2 \,, \left(\hatp_\perp \cdot \r_\perp
+ \r_\perp \cdot \hatp_\perp \right) \right]_{+} \nn \\
& & \quad \quad
+
k \left(z^{\prime \prime}\,, z^\prime \right)
\hatp_{\perp}^2 \hat{L}_z \nn \\
& & \quad \quad
+ A \left(z^{\prime \prime}\,, z^\prime \right)
\left(\hatp_\perp \cdot \r_\perp
+ \r_\perp \cdot \hatp_\perp \right)^{2} \nn \\
& & \quad \quad
+ a \left(z^{\prime \prime}\,, z^\prime \right)
\left(\hatp_\perp \cdot \r_\perp
+ \r_\perp \cdot \hatp_\perp \right) \hat{L}_z \nn \\
& & \quad \quad
+
F \left(z^{\prime \prime}\,, z^\prime \right)
\left(\hatp_{\perp}^2 \r_{\perp}^2
+ \r_{\perp}^2 \hatp_{\perp}^2 \right) \nn \\
& & \quad \quad
+
D \left(z^{\prime \prime}\,, z^\prime \right)
\left[\r_{\perp}^2 \,, \left(\hatp_\perp \cdot \r_\perp
+ \r_\perp \cdot \hatp_\perp \right) \right]_{+} \nn \\
& & \quad \quad
+
d \left(z^{\prime \prime}\,, z^\prime \right)
\r_{\perp}^2 \hat{L}_z \nn \\
& & \quad \quad \left. \left.
+
E \left(z^{\prime \prime}\,, z^\prime \right)
\r_{\perp}^4
\vphantom{\frac{\i}{\LAMBDA}}
\right\} \right]\,.
\eea
The nine aberration coefficients are given by,
\bea
C \left(z^{\prime \prime}\,, z^\prime \right)
& = &
\int_{z^\prime}^{z^{\prime \prime}} d z
\left\{
\frac{1}{8 n_0^3} S^4
- \frac{\alpha_2 (z)}{2 n_0^2} Q^2 S^2
- \alpha_4 (z) Q^4 \right. \nn \\
& & \left. \qquad \qquad
+ \frac{\LAMBDA^2}{2 n_0^3} \alpha_2 (z) \alpha_4 (z) Q^4
\right\} \nn \\
K \left(z^{\prime \prime}\,, z^\prime \right)
& = &
\int_{z^\prime}^{z^{\prime \prime}} d z
\left\{
\frac{1}{8 n_0^3} R S^3
- \frac{\alpha_2 (z)}{4 n_0^2} QS(PS + QR)
- \alpha_4 (z) PQ^3 \right. \nn \\
& & \left. \qquad \qquad
+ \frac{\LAMBDA^2}{2 n_0^3} \alpha_2 (z) \alpha_4 (z) PQ^3
\right\} \nn \\
k \left(z^{\prime \prime}\,, z^\prime \right)
& = &
\frac{\LAMBDA}{2 n_0^3}
\int_{z^\prime}^{z^{\prime \prime}} d z
\alpha_2^2 (z) Q^2 \nn \\
A \left(z^{\prime \prime}\,, z^\prime \right)
& = &
\int_{z^\prime}^{z^{\prime \prime}} d z
\left\{
\frac{1}{8 n_0^3} R^2 S^2
- \frac{\alpha_2 (z)}{2 n_0^2} PQRS
- \alpha_4 (z) P^2 Q^2 \right.  \nn \\
& & \left. \qquad  \qquad
+ \frac{\LAMBDA^2}{2 n_0^3} \alpha_2 (z) \alpha_4 (z) P^2 Q^2
\right\} \nn \\
a \left(z^{\prime \prime}\,, z^\prime \right)
& = &
\frac{\LAMBDA}{2 n_0^3}
\int_{z^\prime}^{z^{\prime \prime}} d z
\alpha_2^2 (z) P Q \nn \\
F \left(z^{\prime \prime}\,, z^\prime \right)
& = &
\int_{z^\prime}^{z^{\prime \prime}} d z
\left\{
\frac{1}{8 n_0^3} R^2 S^2
- \frac{\alpha_2 (z)}{4 n_0^2} (P^2 S^2 + Q^2 R^2)
- \alpha_4 (z) P^2 Q^2 \right. \nn \\
& & \left. \qquad \qquad
+ \frac{\LAMBDA^2}{2 n_0^3} \alpha_2 (z) \alpha_4 (z) P^2 Q^2
\right\} \nn \\
D \left(z^{\prime \prime}\,, z^\prime \right)
& = &
\int_{z^\prime}^{z^{\prime \prime}} d z
\left\{
\frac{1}{8 n_0^3} R^3 S
- \frac{\alpha_2 (z)}{4 n_0^2} PR (PS + QR)
- \alpha_4 (z) P^3 Q \right. \nn \\
& & \left. \qquad \qquad
+ \frac{\LAMBDA^2}{2 n_0^3} \alpha_2 (z) \alpha_4 (z) P^3 Q
\right\} \nn \\
d \left(z^{\prime \prime}\,, z^\prime \right)
& = &
\frac{\LAMBDA}{2 n_0^3}
\int_{z^\prime}^{z^{\prime \prime}} d z
\alpha_2^2 (z) P^2 \nn \\
E \left(z^{\prime \prime}\,, z^\prime \right)
& = &
\int_{z^\prime}^{z^{\prime \prime}} d z
\left\{
\frac{1}{8 n_0^3} R^4
- \frac{\alpha_2 (z)}{2 n_0^2} P^2 R^2
- \alpha_4 (z) P^4 \right. \nn \\
& & \left. \qquad \qquad
+ \frac{\LAMBDA^2}{2 n_0^3} \alpha_2 (z) \alpha_4 (z) P^4
\right\}\,.
\label{fiber-aberration-coefficients}
\eea

Thus we see that the current approach gives rise to all the nine
permissible aberrations.  The six aberrations, familiar from the
traditional prescriptions get modified by the wavelength-dependent
contributions.  The extra three ($k$, $a$ and $d$ are all anisotropic!)
are all pure wavelength-dependent aberrations and totally absent in the
traditional {\em square-root approach}~\cite{DFW,Dragt-Wave} and the
recently developed {\em scalar approach}~\cite{KJS-1,Khan-1}.
A detailed account on the classification of aberrations is available
in~\cite{KBW-1}-\cite{KBW-4}.

\section{Conclusions}
In Part-I and Part-II, we developed an exact matrix representation of
the Maxwell equations which became the basis for an exact formalism
of Maxwell optics.  An exact optical Hamiltonian, with an algebraic
structure in direct correspondence with the Dirac equation of the
electron was derived.  Then following a Foldy-Wouthuysen transformation
technique, a procedure was developed to obtain the beam optical
Hamiltonians to any desired degree of accuracy.  Formal expressions
were obtained for the paraxial and leading order aberrating
Hamiltonians, without making any assumption on the form of the
refractive index.  In this Part we look at the applications of the
above formalism.

First of the two examples is the {\em medium with a constant refractive
index}.  This is perhaps the only problem which can be solved exactly,
in a closed form expression.  This example is primarily for
illustrating certain aspects of the machinery we have used.

The second, and the more interesting example is that of the
{\em axially symmetric graded index medium}.  For this example, in
the traditional approaches one gets only six aberrations.  In our
formalism we get all the nine aberrations permitted by the axial
symmetry.  The six aberration coefficients of the traditional
approaches get modified by the wavelength-dependent contributions.

It is very interesting to note that apart from the wavelength-dependent
modifications of the aberrations, this approach also gives rise to the
image rotation.  This image rotation is proportional to the wavelength
and we have derived an explicit relationship for the angle
in~(\ref{theta}).  Such, an image rotation has no analogue/counterpart
in any of the traditional prescriptions.  It would be worthwhile to
experimentally look for the predicted image rotation.  The existence of
the nine aberrations and image rotation are well-known in {\em axially
symmetric magnetic electron lenses}, even when treated classically.
The quantum treatment of the same system leads to the
wavelength-dependent modifications~\cite{JK2}.

The optical Hamiltonian has two components: {\em Beam-Optics} and
{\em Polarization}.  We have addressed the former in some detail and
shall do the later soon.  The formalism initiated in this article
provides a natural framework for the study of light polarization.
This would provide a unified treatment for the beam-optics and the
polarization.  It also promises a possible generalization of the
{\em substitution} result in~\cite{SSM-2}.  We shall present this
approach soon~\cite{Khan-5}.

The close analogy between geometrical optics and charged-particle
has been known for too long a time.  Until recently it was possible
to see this analogy only between the geometrical optics and classical
prescriptions of charge-particle optics.  A quantum theory of
charged-particle optics was presented in recent
years~\cite{JSSM,J2,CJKP-1,JK2}.  With the  current development of the
non-traditional prescriptions of Helmholtz optics~\cite{KJS-1,Khan-1}
and the matrix formulation of Maxwell optics (in these three Parts),
using the rich algebraic machinery of quantum mechanics it is now
possible to see a parallel of the analogy at each level.
The non-traditional prescription of the Helmholtz optics is in close
analogy with the quantum theory of charged-particles based on the
Klein-Gordon equation.  The matrix formulation of Maxwell optics
presented here is in close analogy with the quantum theory of
charged-particles based on the Dirac equation.  We shall narrate and
examine the parallel of these analogies soon~\cite{Analogy}.

\end{document}